\documentclass[12pt,epsf]{article}
\usepackage{graphicx}

\usepackage{geometry}                
\usepackage{graphicx}
\usepackage{amssymb}
\usepackage{epstopdf}
\usepackage{slashed}
\usepackage{float}
\usepackage{hyperref}
\usepackage{amsmath}
\usepackage{chngcntr}
\usepackage{multirow}
\usepackage{caption}
\usepackage{subcaption}
\numberwithin{equation}{section}

\newcommand{\bea}{\begin{eqnarray}}
\newcommand{\eea}{\end{eqnarray}}
\newcommand{\beas}{\begin{eqnarray*}}
\newcommand{\eeas}{\end{eqnarray*}}
\newcommand{\ba}{\begin{array}}
\newcommand{\ea}{\end{array}}

\newcommand{\pare}[1]{\left( #1 \right)}

\def\be{\begin{equation}}
\def\ee{\end{equation}}
\def\ben{\begin{equation*}}
\def\een{\end{equation*}}
\def\beqa{\begin{eqnarray}}
\def\eeqa{\end{eqnarray}}

\def\be{\begin{equation}}
\def\ee{\end{equation}}

\def\beqa{\begin{eqnarray}}
\def\eeqa{\end{eqnarray}}

\def\bi{\begin{itemize}}
\def\ei{\end{itemize}}

\begin{document}

\begin{titlepage}
\hfill
\vbox{
    \halign{#\hfil         \cr
           } 
      }  
\vspace*{20mm}
\begin{center}
{\Large \bf Extremal Surfaces in Asymptotically AdS Charged Boson Stars Backgrounds}

\vspace*{16mm}
Fernando Nogueira\footnote{nogueira@phas.ubc.ca}
\vspace*{1cm}

{
Department of Physics and Astronomy,
University of British Columbia\\
6224 Agricultural Road,
Vancouver, B.C., V6T 1W9, Canada}

\vspace*{1cm}
\end{center}
\begin{abstract}

In this paper, inspired by the holographic dual of the entanglement entropy, we consider the behaviour of extremal, codimension two, spacelike surfaces in the background of three and four dimensional charged boson stars in asymptotically anti-de Sitter spacetime. We find conditions for which families of minimal area surfaces fail to contain the entire bulk spacetime and construct a phase diagram showcasing the transition between regimes. In addition, we use the relation between the star's mass and the central density of the scalar field to argue for a possible instability of such hollow solutions. Finally, we discuss the consequences of our findings for the study of holographic duals of reduced density matrices.

\end{abstract}

\end{titlepage}

\vskip 1cm


\section{Introduction}


In this paper we construct solutions of asymptotically AdS boson stars coupled to a $U(1)$ gauge field in 3 and 4 dimensions, compute the star's charge and mass as functions of the scalar field central density and study the behaviour of extremal surfaces in these backgrounds. In particular, we determine conditions under which minimal area, codimension 2 spacelike surfaces fail to cover the whole spacetime. To understand the motivation for doing so, let us digress for a moment and consider some intricate open questions regarding the holographic principle in anti de Sitter spacetimes.

The holographic principle is a remarkable idea relating theories with gravity to theories without gravity \cite{thooft,susskind}. The best understood example of holography is the AdS/CFT correspondence which proposes a one to one correspondence between a gravitational theory on anti-de Sitter space and a strongly coupled, large N, conformal field theory living on the conformal boundary of AdS \cite{Maldacena:1997re,Witten:1998qj}. This correspondence has been extensively used to study a variety of strongly coupled field theories and has led to the formulation of successful formalisms such as AdS/QCD and AdS/CMT \footnote{For an extensive review, refer to \cite{McGreevy:2009xe,Hartnoll:2009sz}.}, however, the potential use of the gauge gravity duality to tackle problems in quantum gravity has yet to show comparable progress.

The intrinsic non local aspect of the bulk theory is particularly challenging when trying to understand the exact connection between bulk and boundary degrees of freedom. It stands as one of the major barriers in the way of the gauge / gravity duality fulfilling its potential of answering long standing problems in quantum gravity by recasting them in field-theoretic language. In order to improve this situation, the issue of understanding the map between bulk and boundary degrees of freedom must be addressed. Optimistically, we should be able to express any variation of the boundary theory state in terms of a well defined perturbation of the bulk geometry; conversely, any changes in the bulk should be related to a particular perturbation of the boundary state in a one to one fashion.

An important question related to establishing the dictionary between bulk and boundary is to understand what information about the bulk is contained in a certain region of the boundary field theory. In other words, let a local field theory defined on the conformal boundary $\textbf{B}$ of a spacetime $\mathcal{M}$ be in a generic state $\rho$. Assume this theory has a well defined gravity dual and consider, with respect to some arbitrary time slicing, a spacelike  subregion $A$ of the boundary. 
Which portion of the bulk $R(A)$ is dual to the reduced density matrix $\rho_A$\footnote{Where $\rho_A$ is the partial trace of $\rho$ over the complement of $A$, $\bar{A}$.}, is a question closely related to the issue of characterizing the bulk-boundary degree of freedom map.

The Ryu-Takayanagi proposal \cite{ryu.takayanagi} suggests that the entanglement entropy of a sub-region $A$ of a field theory with a well behaved, static, gravity dual, is proportional to the area of the minimal area surface anchored at the boundary $\delta A$ of $A$. 
Using this proposal we can construct a family of minimal surfaces by considering the surface dual to the entanglement entropy of $\rho_A$ and all the surfaces dual to the partial traces of $\rho_A$ with respect to arbitrary subregions within A. This family of surfaces defines a region in the bulk we will call $w(A)$ and is a possible candidate for $R(A)$ as proposed in \cite{Bartek.et.al1} and further explored in \cite{Wall}\footnote{Formally we should extend the discussion to $D_A$, the causal development of $A$, however, for clarity and objectivity's sake we choose to omit it in the Introduction.}.

Despite meeting several conditions for a suitable candidate for $R(A)$ \cite{Bartek.et.al1,Wall}, when $A=B$, the whole boundary slice, $w({B})$ does not cover the entire spacetime in general. It was pointed out in \cite{Bartek.et.al1} that there are spacetimes for which no minimal surfaces reach certain areas of the bulk that are, nevertheless, causally connected to the boundary, leading us to the conclusion that, at least for such cases, $w({A})$ cannot be the elusive region $R(A)$.

An explicit example of minimal surfaces that do not cover the bulk is found in asymptotically AdS boson star backgrounds. Boson stars are solutions to Einstein's equation coupled to a complex scalar field that have attracted the interest of physicists for over 40 years. Beginning with the work of Kaup \cite{kaup} and others \cite{Ruffini.Bonazzola}, the fields of general relativity, cosmology, and even particle physics have shown great interest in fully understanding these solitonic objects. Some of the standard reviews are \cite{Schunck:2003kk,Jetzer:1991jr,Liebling:2012fv,Liddle:1993ha}.

In this paper we compute the physical charge and mass of asymptotically AdS, charged boson stars as a function of the scalar field central density, investigate numerically the behaviour of extremal surfaces on these backgrounds, and determine the conditions for which $w({A})$ fails to fully cover the bulk, which we will call hollow phases. Furthermore we compare the conditions for the existence of hollow phases with that of known physical instabilities in four dimensional charged boson star systems, and argue that such hollow configurations are likely unstable and therefore physically forbidden, thus providing further evidence that $w(A)$ may be the correct proposal for $R(A)$.

This paper is organized as follows: in section 2 we present the action, the metric and fields ansatz and equations of motion that follow, in addition we discuss boundary conditions and the free parameters we are left with once these are imposed. In section 3 we explore the relation between the star mass, charge, the scalar field central density, and the stability of charged boson stars. We present some numerical results, compare them to what is known from the literature, and argue that for a certain range of parameters the solutions we find are unstable in $D\ge4$ dimensions. In section 4 we review and extend the above discussion as well as outline the numerical strategy to extract information about extremal surfaces and present the reader with preliminary results. In section 5 we construct phase diagrams displaying the relation between the free parameters, the transition between solid and hollow regimes, and the transition between stable and unstable configurations. We conclude with a few final remarks and future directions in section 6.

\section{Equations of Motion}

We start by considering the Einstein-Maxwell action with a negative cosmological constant minimally coupled to a massive complex scalar field in D spacetime dimensions\footnote{Since the main goal of this paper is to discuss numerical solutions of this action, the Hawking-Gibbons surface term can be neglected for it does not alter the equations of motion.}
\be\label{the.action}
S=\frac{1}{8\pi G_D}\int d^Dx\sqrt{-g}\left[\frac{1}{2}R+\frac{(D-2)(D-1)}{2}-\frac{1}{4}F^2-\left|(\partial_{i}-iqA_{i})\Psi\right|^2-m^2 \left|\Psi\right|^2\right].
\ee

We want to restrict our attention to stationary, spherically symmetric configurations and allow for electric charges only, therefore we will let the metric be of the form
\be\label{metric}
ds^2=-f(r)dt^2+\frac{dr^2}{g(r)}+r^2d\Omega_{D-2},
\ee
and adopt the following ansatz for the scalar and gauge fields
\beqa
\Psi&=&\psi(r)\text{e}^{-i \omega t},\label{psioft}\\
A_0&=&A(r),\\
A_{i\neq 0}&=&0.
\eeqa

To include the gravitational contribution from both the scalar and gauge fields we write the total energy momentum tensor as
\be
T_{ij}=T^{SF}_{ij}+T^{GF}_{ij},
\ee
the sum of the scalar's energy momentum tensor
\be
T^{SF}_{ij}=\nabla_i\Psi^*\nabla_j\Psi+\nabla_i\Psi\nabla_j\Psi^*-g_{ij}\pare{\left|\nabla\Psi\right|^2+m^2\left|\Psi\right|^2},
\ee
where $\nabla_i$ is the covariant derivative $(\partial_{i}-iqA_{i})$, and the gauge field energy momentum tensor
\be
T^{GF}_{ij}=F_{ik}F^k_{j}-\frac{1}{4}g_{ij}F^2.
\ee
Additionally, note that the action \ref{the.action} is invariant under the global $U(1)$ rotation $\Psi \rightarrow \text{e}^{i \alpha}\Psi$, implying the existence of the conserved current
\be
J^j=i g^{jk}((\nabla_k\Psi)^*\Psi-(\nabla_k\Psi)\Psi^*),
\ee
which will act as a source for the gauge field.

With the above definitions in hand we obtain four linearly independent equations of motion from the {\it tt} and {\it rr} components of Einstein's equation
\be\label{Eeq}
R_{ij}-\frac{1}{2}g_{ij}R-\frac{(D-2)(D-1)}{2}g_{ij}=T_{ij},
\ee
the Klein-Gordon equation
\be
\nabla^2\phi - m^2 \phi=0,
\ee
and Maxwell's equations
\be
\nabla_iF^{ij}=qJ^j.
\ee


In terms of the functions defined in our ansatz, the D dimensional equations of motion from the tt and rr components of Einstein's equation are
\beqa\label{eom1}
-q^2 r^2 A(r)^2 \psi (r)^2-2 q r^2 \omega  A(r) \psi (r)^2-\frac{1}{2} (D-2) r^{D-3} f(r)g'(r)-&&\nonumber\\
\frac{1}{2} (D-3) (D-2) f(r) g(r)+\frac{1}{2} (D-2) (D-1) r^2 f(r)-r^2 g(r) A'(r)^2+&&\nonumber\\
\frac{1}{2} (D-3) (D-2) f(r)-r^2 f(r)g(r) \psi '(r)^2-m^2 r^2 f(r) \psi (r)^2-r^2\omega^2\psi (r)^2&=&0,\nonumber\\
&&
\eeqa
and
\beqa\label{eom2}
q^2 r^2 A(r)^2 \psi (r)^2+2 q r^2 \omega  A(r) \psi (r)^2-\frac{1}{2} (D-2) r^{D-3} f'(r)g(r)-&&\nonumber\\
\frac{1}{2} (D-3) (D-2) f(r) g(r)+\frac{1}{2} (D-2) (D-1) r^2 f(r)-r^2 g(r) A'(r)^2+&&\nonumber\\
\frac{1}{2} (D-3) (D-2) f(r)+r^2 f(r)g(r) \psi '(r)^2-m^2 r^2 f(r) \psi (r)^2+r^2 \omega ^2 \psi (r)^2&=&0.\nonumber\\
&&
\eeqa
While from the Klein-Gordon equation we find
\beqa\label{eom3}
2 q^2 r A(r)^2 \psi (r)+2 (D-2) f(r) g(r) \psi '(r)+r g(r) f'(r) \psi '(r)+&&\nonumber\\
r f(r) g'(r) \psi '(r)+2 r f(r) g(r)\psi ''(r)-2 m^2 r f(r) \psi (r)+2 r \omega ^2 \psi (r)&=&0,\nonumber\\
&&
\eeqa
and Maxwell's equations give
\beqa\label{eom4}
-2 r f(r) g(r) A''(r)-2 (D-2) f(r) g(r) A'(r)+r g(r) A'(r) f'(r)-&&\nonumber\\
r f(r) A'(r) g'(r)+4 q^2 r A(r) f(r) \psi (r)^2+4q r \omega  f(r) \psi (r)^2&=&0.\nonumber\\
&&
\eeqa

These comprise a set of two first order and two second order ordinary differential equations and allow for a six parameter family of solutions, one of which is empty AdS. However, we are not interested in any generic solution of the above set of equations, but only in those that are asymptotically AdS and regular at $r=0$. 

By requiring the solution to be regular at the origin, the near $r=0$ analysis of the equations of motion leads to the following constraints
\be
g(0)=1,\quad g^{\prime}(0)=0, \quad f(0)=f_0,\quad f^{\prime}(0)=0,
\ee
\be
A(0)=A_0,\quad A^{\prime}(0)=0, \quad \psi(0)=\psi_0,\quad \psi^{\prime}(0)=0,
\ee
leaving three undetermined parameters. We can use the following symmetry of the equations of motion
\be
f\rightarrow \gamma^2 f,\quad A\rightarrow \gamma A,\quad \omega\rightarrow\gamma\omega,\quad t\rightarrow\frac{1}{\gamma}t,
\ee
to fix $A(0)=1$. In addition, the value of $f_0$ is chosen to ensure that our solutions asymptote to global AdS, that is as $r\rightarrow \infty$, $f(r)\rightarrow 1+r^2$, and similarly for $g(r)$. This leaves us with one free parameter, $\psi_0$, and the task of studying a one parameter family of solutions of equations \ref{eom1}-\ref{eom4}.

However, before we move forward, we are still left with the issue of guaranteeing that the scalar field showcases the correct asymptotic fall off. In general, the large $r$ behaviour of $\psi(r)$ is given by

\be\label{scalarinfinity}
\psi(r)=\frac{\psi_1}{r^{\lambda_+}}+\frac{\psi_2}{r^{\lambda_-}},
\ee
with
\be
\lambda_{\pm}=\frac{1}{2}\pare{(D-1)\pm\sqrt{(D-1)^2+4m^2}},
\ee
where $m^2$ is the scalar field mass in the lagrangian \ref{the.action}, and $\psi_1$ and $\psi_2$ are constants. For $m^2>0$, $\psi(r)$ has a non normalizable term that can be set to zero by picking a solution for which $\psi_1=0$, while for $(D-1)^2/4<m^2<0$ both terms are normalizable and therefore allowed in principle.

To pick the desired fall off of $\psi(r)$ we will look for the lowest value of $\omega$, the phase of $\Psi(t)$, for which we observe $\psi_1=0$. We do so by using $\omega$ as a shooting parameter and imposing the boundary condition $\psi_1< 10^{-10}$. Once $\omega$ is fixed we are left with only $m^2$ and $q$ (the scalar charge) as free theory parameters. Henceforth in this paper we will numerically study the one parameter family of solutions of equations \ref{eom1}-\ref{eom4}, their dependence on the solution's parameter $\psi(0)$, the central density of the scalar field, as well as on the theory's parameters $m^2$ and $q$.

\section{Mass, Charge and Scalar Central Density}

Before we start the discussion of how the solutions of equations \ref{eom1}-\ref{eom4} and the families of minimal surfaces depend of the free parameters discussed above, lets take some time to look at how the total mass and total charge of the star are calculated and how they depend on the central density of the scalar field, $\psi(0)$. 

The metric (\ref{metric}) describes a spherically symmetric body whose mass can be extracted from the asymptotic behaviour of the metric function $g(r)$. At large r we expect the metric to approach that of a charged, massive star in AdS with no scalar field, i.e.:
\be\label{asympmetric}
g(r)\xrightarrow{r\rightarrow\infty}1+r^2-\frac{2M}{r^{D-3}}+\frac{Q^2}{r^{2(D-2)}},
\ee
where $M$ is the mass of the star and $Q$ its charge, which is simply the total number of scalar particles times the charge $q$, i.e.:
\be\label{TheCharge}
Q=q\int d^{D-1}x J^0\sqrt{-g}.
\ee

However, the existence of a scalar field allows for the possibility of scalar hair, therefore we should expect that our metric, while still approaching \ref{asympmetric} at large r, will receive higher order corrections. Ergo we choose to write the general metric as
\be
g(r)=1+r^2-\frac{2M(r)}{r^{D-3}}+\frac{Q(r)^2}{r^{2(D-2)}},
\ee
where the large r behaviour of the function $M(r)$ is given by
\be\label{TheMass}
M(r)\sim M + \mathcal{O}\pare{1/r^{\alpha}},
\ee
with $\alpha$ being is a positive constant, and similarly for $Q(r)$.

The behaviour of the mass and charge of a boson star as a function of the scalar's central density has been extensively studied in the literature both for asymptotically flat and asymptotically AdS spacetimes \cite{Astefanesei.Radu,CBS-phaseT}. In particular, in AdS space, the existence of a variety of new solutions and the presence of a zero temperature phase transition have recently been shown to exist \cite{CBS-phaseT,Gentle:2011kv}.  Furthermore the stability of boson stars has also been the subject of numerous studies and found to correlate with certain aspects of the behaviour between mass, charge and central density \cite{Astefanesei.Radu,Jetzer:1989us} as we discuss below. 

Since later in this paper the question of whether the solutions we find displaying hollow configurations of minimal area surfaces are physically stable will be particularly important, we should take a closer look at the behaviour of the mass and charge of our solutions as a function of the central density $\psi(0)$ and compare it to what is known from the literature. Our analysis here will rely solely on previously known results and educated guesses, so we will refrain from a formal study of the stability of the solutions as it lies outside the scope of this work.  

For $D\ge 4$ dimensions it has been found that for models with no gauge field in both zero and negative cosmological constants background, and with a gauge field in a zero cosmological constant background, the star mass as a function of the scalar field central density $M(\psi(0))$ reaches a maximum value for a finite $\psi(0)=\psi_c$. Even though the numerical value for the maximum mass and $\psi_c$ change for each case, they correspond to the threshold between dynamically stable ($\psi(0)<\psi_c$) and unstable ($\psi(0)>\psi_c$) regimes\footnote{At maximum mass ($\psi(0)=\psi_c$) the pulsation equation arising from the analysis of the time evolution of infinitesimal radial perturbations has a zero mode indicating that $\psi_c$ is a boundary between stable and unstable equilibrium configurations \cite{Jetzer:1989us}.}.

The $D=3$ case has attracted considerably less attention and consequently, to the extent of the author's knowledge, no formal result is available. Nevertheless, in the context of strongly self interacting boson stars, there is evidence for the existence of a maximum mass as well as some discussion regarding the positive binding energy of these objects possibly being an indicator of instabilities \cite{Sakamoto:1998hq,Sakamoto:1998aj}. Despite the existence of such partial results we will refrain from making any statements regarding the stability of the three dimensional solutions.

\begin{figure}[tb!]
\centering%
\includegraphics[scale=1.2]{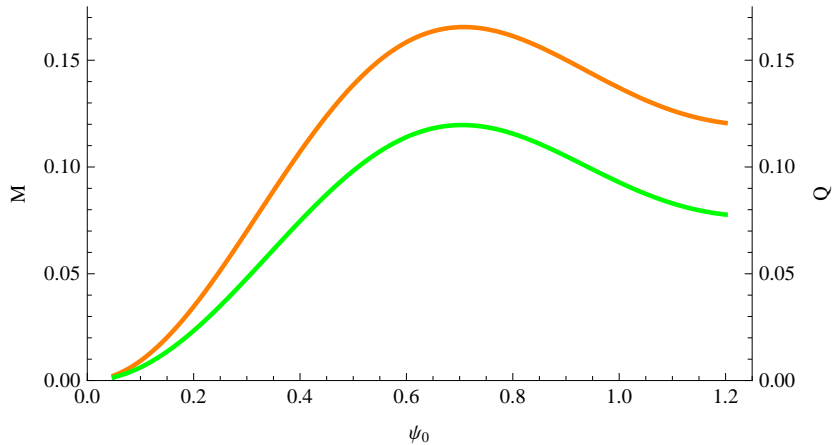}
\caption{Plot of the star's mass (orange) and charge (green) versus the central value of the scalar field $\psi_0$ with $m^2=0$ and $q=0.2$ in $D=4$ dimensions.}
\label{MaxMQ4D}
\end{figure}

With the above discussion in mind we numerically solve equations \ref{eom1}-\ref{eom4}, compute the physical mass and charge of the configurations using \ref{TheMass} and \ref{TheCharge} for different values of $\psi(0)$ and compare to what is known from the literature.  We find qualitatively similar results in $D=4$ to that of pure boson stars (without gauge field), more importantly we observe the existence of a maximum value for $M$ as can be seen in figure \ref{MaxMQ4D}. 

Given the proximity of the models considered in the literature and ours, together with the qualitative agreement between results, we are led to believe that our model is also unstable past the central density threshold $\psi_c$ for $D\ge 4$ dimensions. We will assume this conjecture is indeed true in the reminder of this paper and use it to understand better the conditions for the existence of a hollow $w(D_A)$ in the upcoming sections.

\begin{figure}[bt!]
\centering%
\includegraphics[scale=1.2]{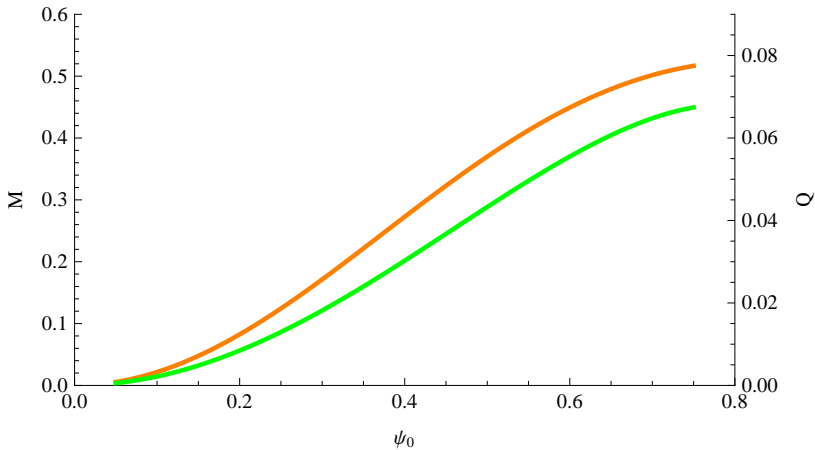}
\caption{Plot of the star's mass (orange) and charge (green) versus the central value of the scalar field $\psi_0$ with $m^2=0$ and $q=0.2$ in $D=3$ dimensions.}
\label{MaxMQ3D}
\end{figure}

In $D=3$ dimensions we again observe qualitatively similar results to those found in \cite{Astefanesei.Radu} (figure \ref{MaxMQ3D}). Most notably we would like to highlight the absence of a maximum mass within the central density range investigated\footnote{Our numerical solution is untrustworthy past the $\psi(0)=0.8$ mark in $D=3$ dimensions, therefore precluding us from investigating the mass versus central density relation any further.} as it showcases a considerably different behaviour than that expected for $D\ge 4$ dimensions.

\section{Overlapping Extremal Surfaces}

In the Introduction we raised the question of what portion $R(A)$ of the bulk spacetime $\mathcal{M}$ is dual to the reduced density matrix $\rho_A$ for the causal development $D_A$ of a given region $A$ of the boundary theory. 
A natural starting point would be to consider the causal wedge $z(D_A)=J^+(D_A)\cap J^-(D_A)$\footnote{Where $J^+(D_A)$, defined as all points accessible by causal curves arising from $D_A$, is the causal future of $D_A$, while $J^-(D_A)$, defined as the set of points from which $D_A$ can be reached following a causal curve, is its causal past.}, the region in the bulk a boundary observer restricted to $D_A$ is causally connected to. Intuitively speaking, we expect $z(D_A)$ to be at least the minimal portion of the bulk accessible to a boundary observer living in $D_A$. The reason being that its complete causal connection to $D_A$ allows for the detection of any perturbation of the bulk metric inside $z(D_A)$.

While $z(D_A)$ seems to be a natural guess, and it certainly imposes a lower bound on the size of $R(A)$, examples for which $z(D_A) \subset w(D_A)$ while $z(D_A)\nsupseteq w(D_A)$, are easy to find\footnote{See \cite{Bartek.et.al1} for a thorough discussion.} and demonstrate how the region $z(D_A)$ alone cannot, in general, hold all the information stored in $D_A$. In contrast, for some cases, we can argue that all the information in $D_A$ must lie within $\hat{w}(D_A)$, the domain of dependence of $w(D_A)$, for anything outside this bulk region will causally interact with the complement of $D_A$, $D_{\bar{A}}$. This could lead us to naively expect that $\hat{w}(D_A)$ imposes an upper bound on $R(A)$, however, one can construct explicit examples for which this might not be true, as we will see below.

Say that we let the region $D_A$ cover the entire boundary, i.e.: $D_A=\textbf{B}$, an observer within $D_A$ will have access to a full Cauchy surface and, consequently, information about the entire past and future of the boundary theory. Since we are considering a field theory with a well defined gravity dual, information about the past and future of the boundary should extend to information about the past and future of the bulk theory as well. In other words, an observer in $\textbf{B}$ that has access to all possible boundary physical observables should be able to fully reconstruct the dual bulk metric\footnote{This is demonstrably true for the case of empty AdS \cite{Hamilton:2005ju,Hamilton:2006az,Heemskerk:2012mn,Kabat:2011rz}. Some interesting recent discussion highlights the necessity of including non local boundary operator in the boundary observer's toolbox \cite{Bousso:2012mh}}.

If we restrict the access of this observer to knowledge of the entanglement entropy for any arbitrary region within $\textbf{B}$, we can ask how much of the bulk he or she can probe and, better yet, infer the geometry of \cite{Bartek.et.al1,Wall,Czech:2012be}. If the bulk theory has a horizon, say a spherically symmetric black hole at the origin, it was shown in \cite{Hubeny} that, while no extremal surface of any co-dimension (or causal lines, for that matter) can probe inside the horizon, given a fixed boundary region, co-dimension 2 surfaces probe the bulk deeper than higher co-dimension surfaces or causal lines. Now, if instead of a black hole at the origin we consider a boson star for example, we can ask the same question again: how much of the bulk spacetime can a boundary observer probe with extremal surfaces only? Note that now the entire bulk is causally connected to the boundary, so we know that $z(\mbox{{\bf B}}) \supset \mathcal{M}$, therefore, if $w(D_{{\bf B}})$ fails to fully cover the bulk, we will have an explicit example for which $z \supset w$ while $w\nsupseteq z$.

Given the above discussion, our goal is to search within the space of solutions of charged boson stars in asymptotically AdS spacetime for configurations for which we observe overlapping of extremal surfaces leading to a hollow $w(D_A)$, therefore addressing the question of whether $z(D_A) \supset w(D_A)$ while $w(D_A)\nsupseteq z(D_A)$ is feasible\footnote{Note that failure to find solutions obeying $z(D_A) \supset w(D_A)$ while $w(D_A)\nsupseteq z(D_A)$ does not indicate this particular phenomena is not possible, in the same way that the mere mathematical existence of such configurations is not enough to undermine the candidacy of $w(D_A)$ to the position of $R(A)$ for these could not be physically preferred.}.

\begin{figure}[htb!]
\centering%
\includegraphics[scale=.75]{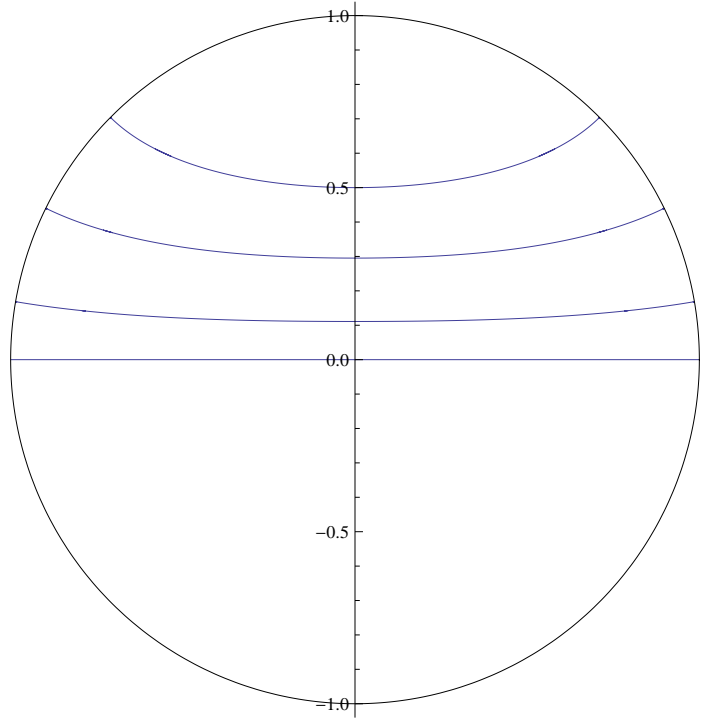}
\caption{A plot of the Penrose diagram of a time slice of multiple extremal surfaces on an asymptotically AdS charged boson boson star background in global coordinates for $m^2=0,$ $q=0.1$, and $\psi(0)=0.2$. From top to bottom we have $\theta=\pm0.251\pi, \pm0.355\pi, \pm0.446\pi, \pm0.5\pi$. In this particular case the central density of the scalar field is below the threshold $\psi_h$, therefore there are no degenerate extremal surfaces (see figure \ref{geodesics}), in other words, we observe a solid $w(D_A)$.}
\label{sts}
\end{figure}

\begin{figure}[htb!]
\centering%
\includegraphics[scale=.75]{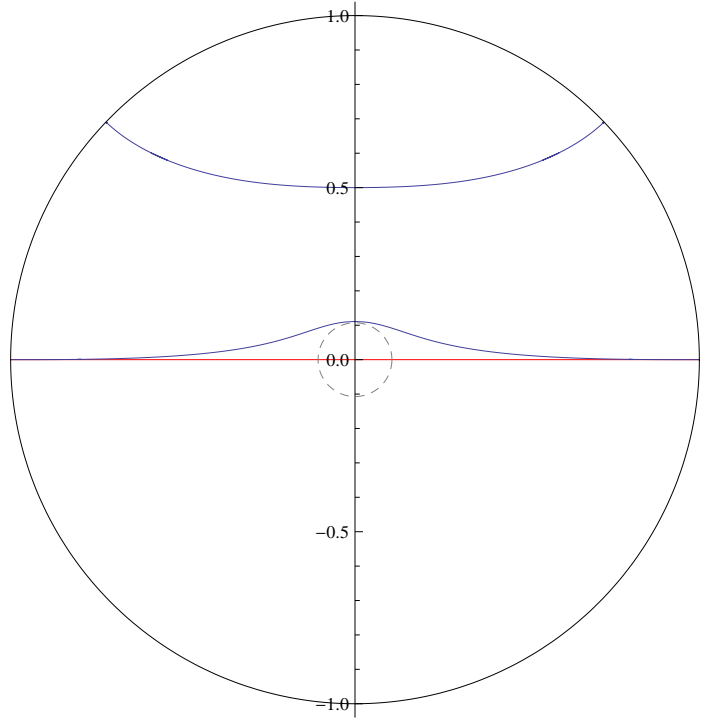}
\caption{Again, a plot of the Penrose diagram of a time slice of multiple extremal surfaces on an asymptotically AdS charged boson boson star background in global coordinates for $m^2=0,$ $q=0.1$, and $\psi(0)=1.2$. However, in this example, the scalar central density is above the threshold $\psi_h$ and we observe the existence of degenerate extremal surfaces for a range of boundary anchor points $\theta$ (see figure \ref{geodesics}). In particular, for anchor points $\theta=\pm \pi/2$ there are three solutions two of which (blue line) lie on top of each other, have minimal area and do not penetrate the dashed small circle, while the third (red line) corresponds to a non minimal area extremal surface. As discussed in this paper, no minimal area surface penetrates the deepest bulk points within the small dashed circle.}
\label{hts}
\end{figure}

\begin{figure}[htb!]
\centering%
\includegraphics[scale=.75]{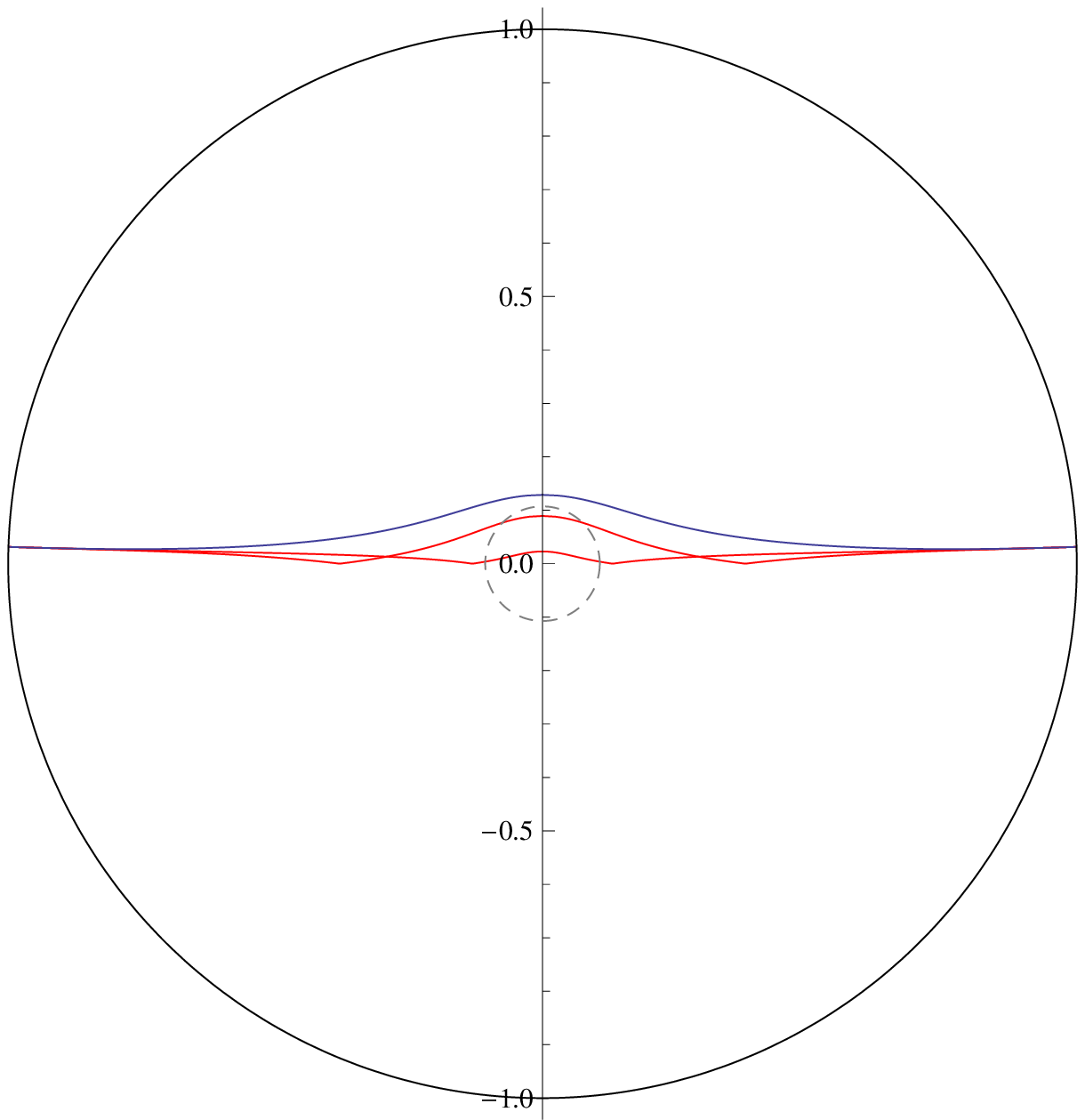}
\caption{A similar plot as figures \ref{sts} and \ref{hts} highlighting the behaviour of all three extremal surfaces anchored at the same boundary points $\theta = \pm 0.49 \pi$. Although two of the solutions (red lines) penetrate the dashed circle, these are not the minimal area and therefore are not part of the $w(D_A)$ set.}
\label{h49ts}
\end{figure}

Our setup is both static and asymptotically globally AdS, therefore our boundary is a sphere. We will look for surfaces that extend in all polar angles, so, to our applications, it suffice to describe them as curves $\theta(r)$, with anchor points $\theta(\infty)=\pm\theta_0$ corresponding to the azimuthal boundary coordinates of the start and end points (see figures \ref{sts}, \ref{hts} and \ref{h49ts}). 

To determine whether the entangling surfaces reach the deepest regions in the bulk we will analyze the behaviour of $r_{\mathrm{min}}(\theta_n(r))$ for multiple surfaces with distinct boundary anchor points $\theta_n(\infty)=\pm\theta_n$, that is, the minimum value of $r$ reached by a given entangling surface $\theta_n(r)$ with fixed boundary anchor points $\theta_n$. 

\begin{figure}[htb!]
\hspace{-0.6cm}
        \centering
        \begin{subfigure}[b]{0.55\textwidth}
                \centering
                \includegraphics[scale=.55]{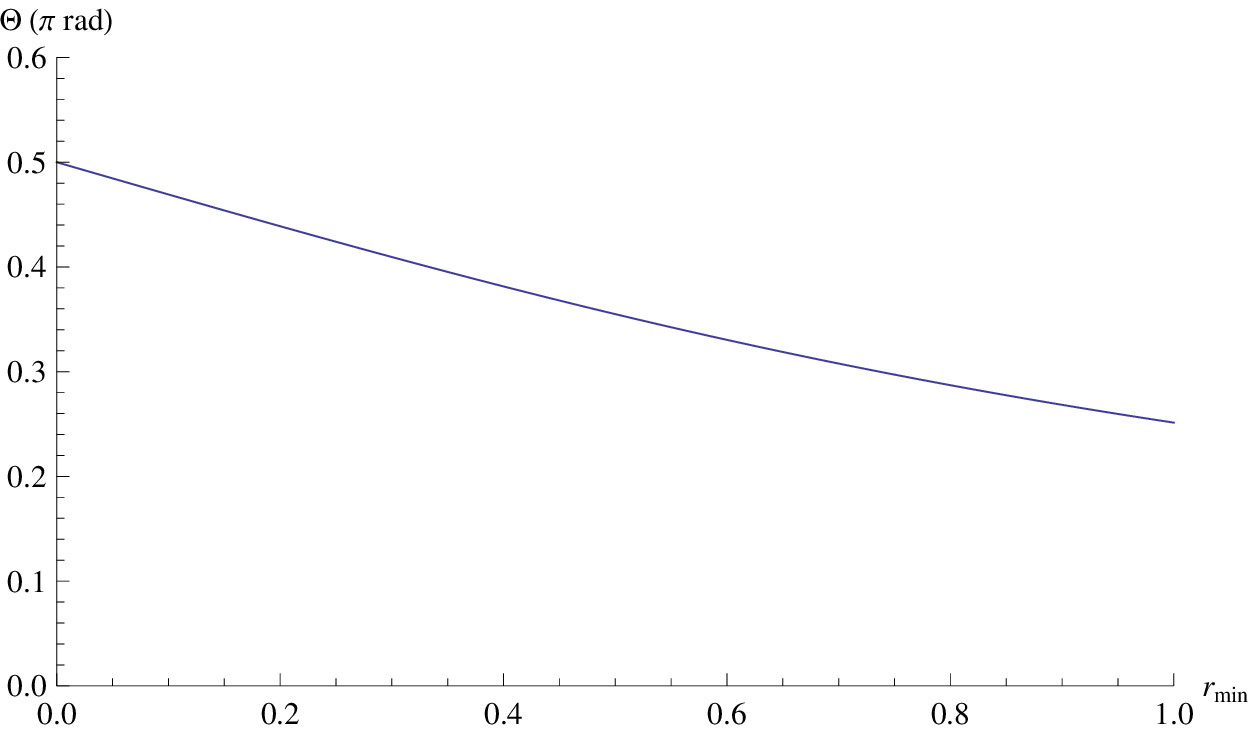}
                \caption{$\psi(0)=0.2$}
                \label{geodesics1}
        \end{subfigure}%
\hspace{-1.18cm}
        \begin{subfigure}[b]{0.55\textwidth}
                \centering
                \includegraphics[scale=.55]{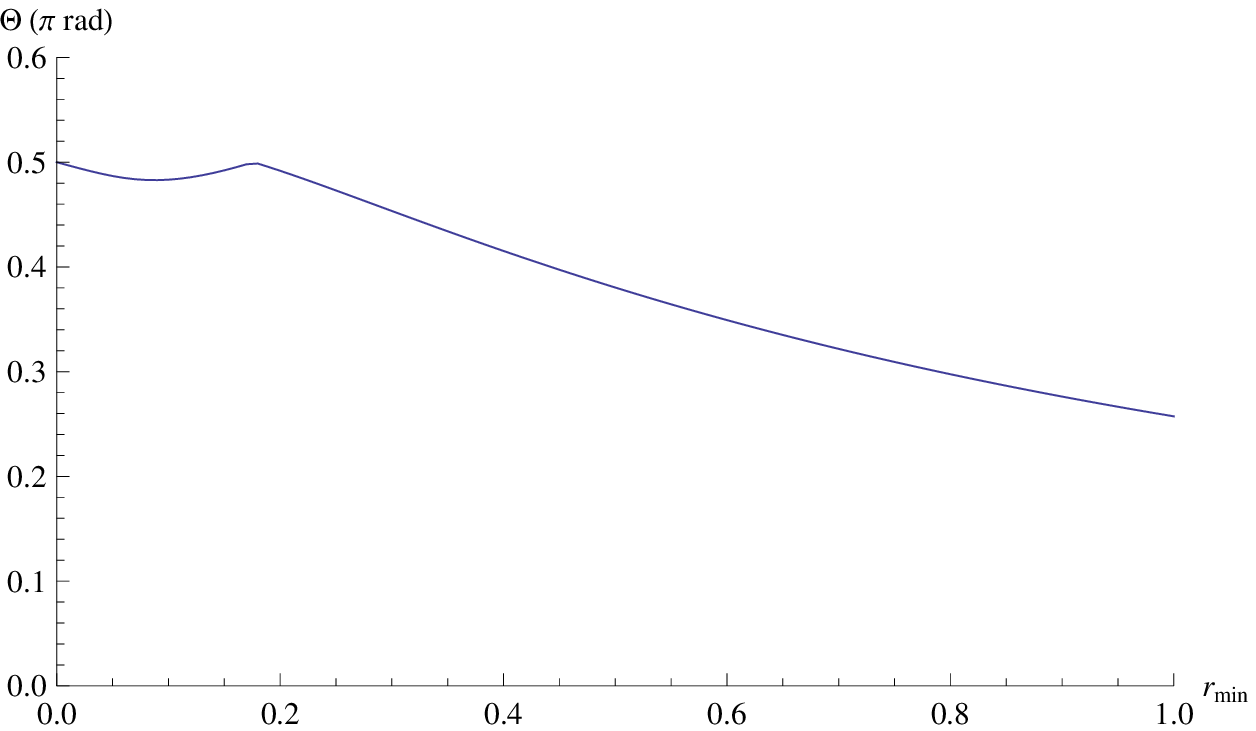}
                \caption{$\psi(0)=1.2$}
                \label{geodesics2}
        \end{subfigure}
        \caption{A comparison of two different central densities for the plot of  $r_{\mathrm{min}}(\theta(r))$ in four dimensions, both cases with $m^2=0$ and $q=0.1$. It is clear how for anchor points roughly between $0.17\pi<\theta<\pi/2$ there exist three distinct extremal surfaces with different values of $r_{\mathrm{min}}$, however, only one of them has minimal area. See figures \ref{sts}, \ref{hts} and \ref{h49ts} for specific examples.}
\label{geodesics}
\end{figure}

We know that $\theta(\infty)_{r_{\mathrm{min}}}$ (the anchor points of a surface that has $r_{\mathrm{min}}$ as its deepest point) covers all values of $\theta$ as we vary $r_{\mathrm{min}}$ from zero to infinity, as a result, for every $0<r_0<\infty$ there is at least one extremal surface that obeys $r_{\mathrm{min}}(\theta(r))=r_0$ (figure \ref{geodesics}).

However, the situation is much different if we focus on surfaces anchored at the same boundary points. In this case we expect that only one minimal area surface can be anchored at each given $\theta_i$ (figure \ref{geodesics1}). While this is true in general (with the exception of $\theta=\pi/2$), this need not be the case if we consider extremal rather than minimal area surfaces (figure \ref{geodesics2}), therefore we should not be alarmed if for certain boundary anchor points we find that there are multiple distinct extremum surfaces anchored to it, in fact, we are most interested in determining when such a phenomena happens. To do so we will numerically investigate conditions under which there exist multiple extremal surfaces anchored at the same boundary point that do not share the same deepest point in the bulk.

The equation describing extremal surfaces can be found by minimizing the area of a spacelike codimension 2 surface given by
\be
A=\mbox{Vol}(S^{D-3})\int{dr \pare{\sin{\theta(r)}}^{D-3}  r^{D-3} \sqrt{\frac{1}{g(r)}+r^2\pare{\frac{d\theta}{dr}}^2}},
\ee
leading to a second order differential equation for the function $\theta(r)$ which can be solved analytically in $D=3$ dimensions or numerically for $D\ge 4$ dimensions. In any case, knowledge of $\theta(r)$ allow us to search for different solutions $\theta_i(r)$ for which $\theta_i(\infty)=\theta_0,\quad \forall i,$ while $r_{\mathrm{min}}(\theta_i(r))\neq r_{\mathrm{min}}(\theta_{i^{\prime}}(r))$ for at least one $i^{\prime}$. In other words, it allow us to look for extremal surfaces anchored at the same boundary points that do not share a common deepest bulk point $r_{\mathrm{min}}(\theta_i(r))$ (figures \ref{hts} and \ref{h49ts}).

The existence of such multiplicity of extremal surfaces with common boundary anchor points will in general preclude most of them from having minimal area and, as argued in \cite{Bartek.et.al1}, can be used to show that no family of minimal area surfaces can cover the bulk in its entirety. Nevertheless, as a consistency check, we computed numerically the area for surfaces anchored at various boundary points and confirmed that, indeed, when more than one solution exists (with the same $\theta(\infty)$), the one with the bigger $r_{\mathrm{min}}$ has the smallest area. 

Therefore, our strategy to determine whether our charged boson star solutions display such hollow phases is to look for an extremal surface $\theta_h(r)$ that obeys $\theta_h(\infty)=\pi/2$ while having $r_{\mathrm{min}}(\theta_h)\neq 0$ (figure \ref{hts}) since, from the spacetime symmetry, we know that there exist one extremal surface for which $\theta(\infty)=\pi/2$ with $r_{\mathrm{min}}(\theta)= 0.$

As a warming up exercise we fix the value of the parameters $m^2$ and $q$ and, as we vary the central density of the scalar field $\psi_0$, we observe the system to transition between solid and hollow phases as seen in figure \ref{geodesics}. Note how on figure \ref{geodesics1} there is only one solution extending into the bulk and reaching a specific $r_{\mathrm{min}}$ for each $\theta$, whereas on figure \ref{geodesics2} there is more than one value of $r_{\mathrm{min}}$ for a given boundary anchor point (a fixed $\theta_0$) near the $r=0$ region.

Our goal for the next section is to determine the precise value $\psi_h$ of $\psi(0)$ for which this transition occur as a function of the parameters $m^2$ and $q$ and, in the four dimensional case, compare it to $\psi_c$, the central density threshold between stable and unstable configurations.

\section{Phase Diagrams}

In this section we explore in further detail the conditions for which we observe solid and hollow phases. Since we are dealing with a one parameter family of solutions and have two free theory parameters, we should be able to construct a three dimensional phase diagram and find a two dimensional surface separating solid and hollow phases. To numerically accomplish this task we start by fixing $q$ while varying $m^2$. For each value of $m^2$ and $q$ we look for the lowest value of $\psi(0)$ for which we observe multiple values of $r_{\mathrm{min}}(\theta_0)$ for the same, fixed, $\theta_0$ and find a line separating the two regions, we then repeat this process for multiple different values of $q$. 

\begin{figure}[htb!]
\centering%
\includegraphics[scale=1]{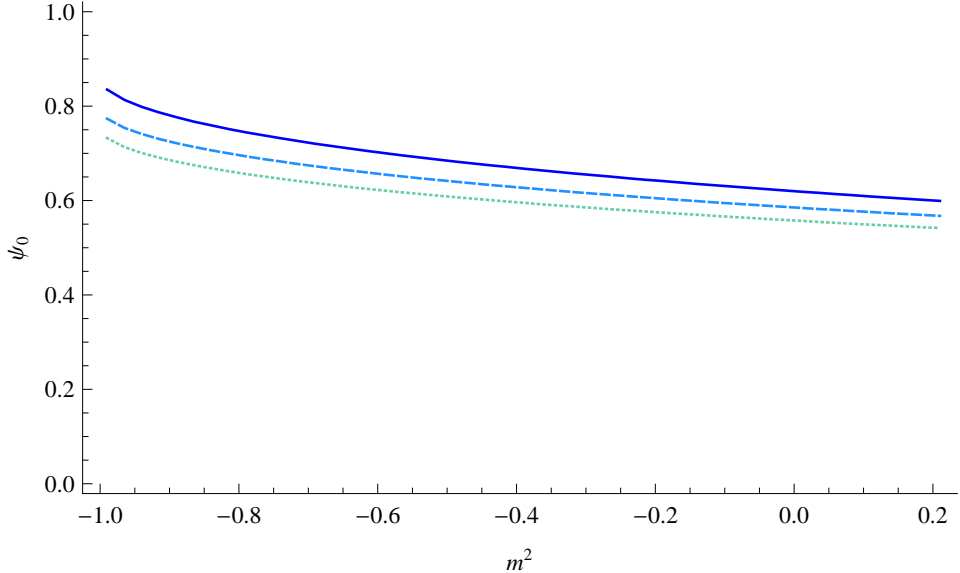}
\caption{Critical scalar field central density separating solid and hollow configurations for $D=3$ dimensions with $q=0.1$ in blue, $q=0.2$ in light blue (dashed) and $q=0.3$ in green (dotted). In all cases a central density value below the line correspond to solid solutions, while above it lies the hollow regime.}
\label{SH3D}
\end{figure}

\begin{figure}[htb!]
\centering%
\includegraphics[scale=1]{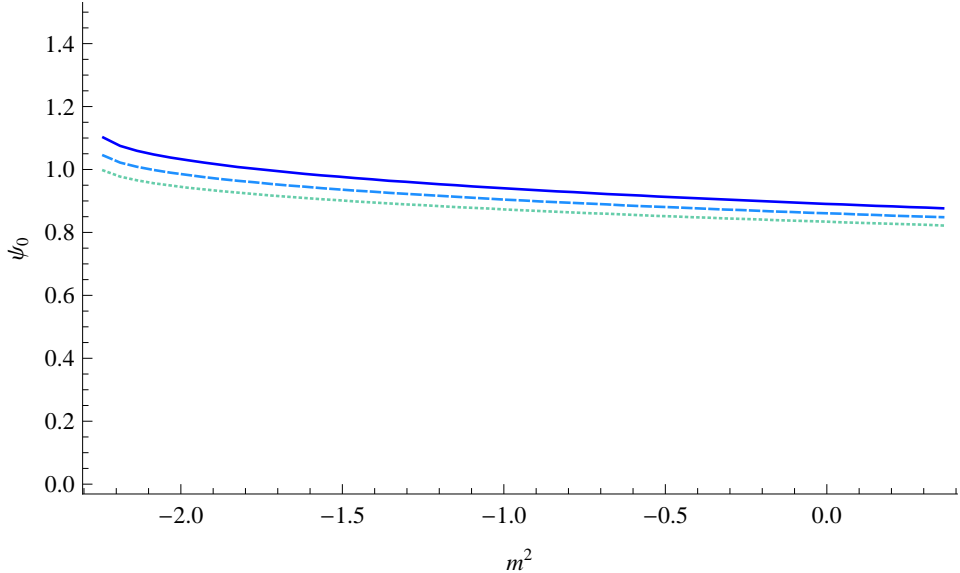}
\caption{Critical scalar field central density separating solid and hollow configurations for $D=4$ dimensions with $q=0.1$ in blue, $q=0.2$ in light blue (dashed) and $q=0.3$ in green (dotted). In all cases a central density value below the line correspond to solid solutions, while above it lies the hollow regime.}
\label{SH4D}
\end{figure}

As seen in figures \ref{SH3D} and \ref{SH4D}, we observe a similar behaviour for both three and four dimensional cases. Is is clear that the threshold value of $\psi(0)$ decreases as we increase either $m^2$ or $q$, and a maximum, finite value is attained as the scalar mass approaches the BF bound and the charge goes to zero.

As discussed earlier in this paper, four dimensional boson stars are known to be unstable when the central scalar density rises above a critical value $\psi_c$ in numerous different setups. Since we just established the existence of $\psi_h$, the threshold between solid and hollow configurations, we want to compare it to $\psi_c$ so we can determine whether the hollow solutions we find are in fact physically permitted.

\subsubsection*{Stability of AdS charged boson stars and hollowed phases.}

The stability of boson stars has been a subject of intensive study in the past decades, while focus has been given to boson stars in flat spaces, similar results exist in AdS space. Despite subtle changes between the flat, curved, self interacting or charged cases, it is well known that four dimensional boson stars reach a maximum mass value for a finite central density $\psi_c$ and are unstable past this point. The nature of the instability and how it depends on the various variations of the model, while important on their own, are not within the scope of this study, for us it suffices to know that the value of $\psi(0)=\psi_c$ for which the stars mass as a function of scalar central density, $M(\psi(0))$, is maximum represents the threshold between stable and unstable regimes and that this seems to be universal across different types of boson stars\footnote{For the interested reader we direct you the reviews \cite{Schunck:2003kk} and \cite{Liebling:2012fv} and the work \cite{Astefanesei.Radu} for a lengthy discussion on boson star instabilities and more.}.

\begin{figure}[htb!]
\centering%
\includegraphics[scale=1]{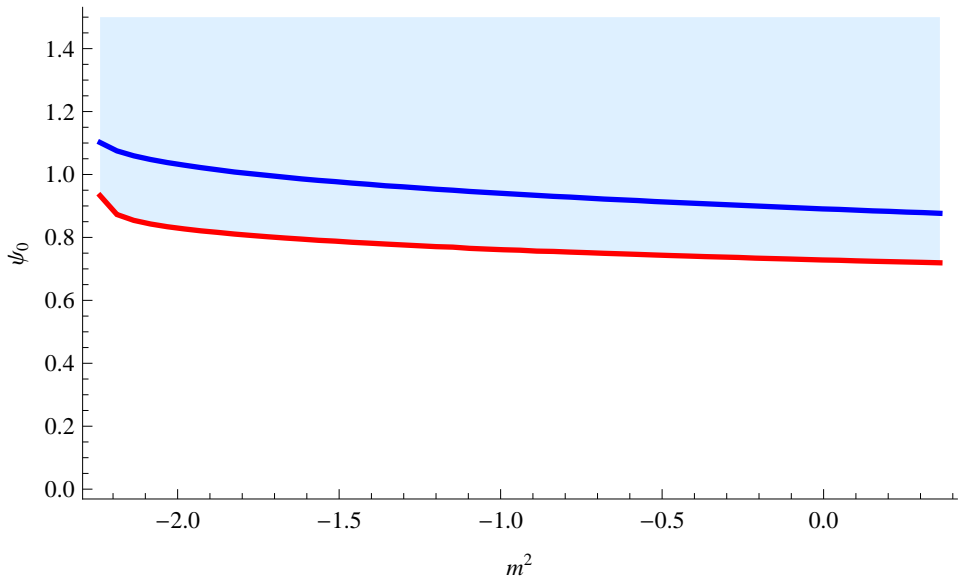}
\caption{A phase diagram displaying both transitions found in $D=4$ dimensions (stable $\rightarrow$ unstable and solid $\rightarrow$ hollow) with $q=0.1$. The red line (below) is the stability threshold, a value of $\psi_0$ above it (the light blue region) renders a dynamically unstable configuration. The blue line (above), once again, represents the transition between solid and hollow configurations. It is clear from this figure how, in the range studied, only solid configurations are physically allowed.}
\label{UnstablexHollow}
\end{figure}

\begin{figure}[htb!]
\centering%
\includegraphics[scale=1]{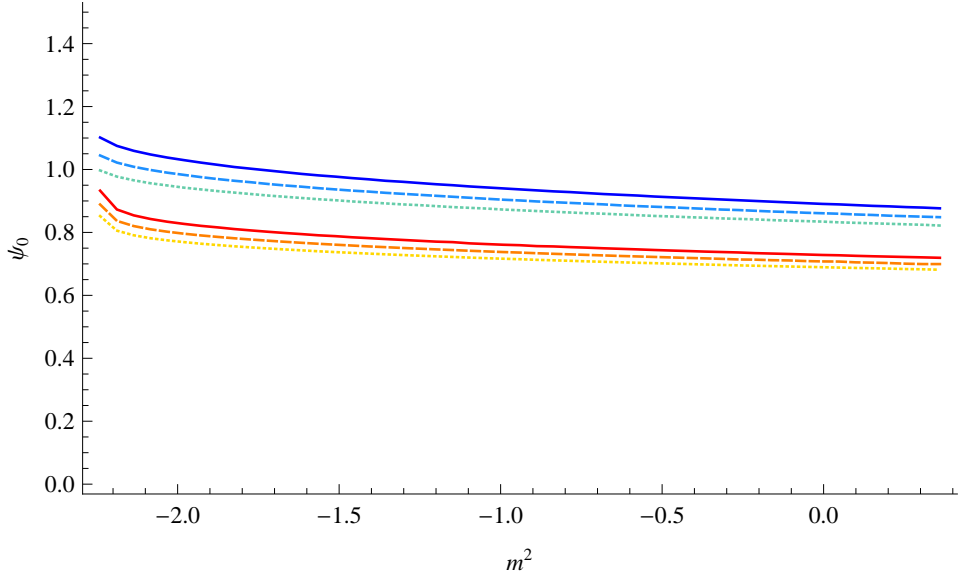}
\caption{A phase diagram comparing the central densities $\psi_c$ and $\psi_h$ as a function of $m^2$ for different values of $q$. The dotted lines are for $q=0.3$, the dashed lines for $q=0.2$, while the solid lines for $q=0.1$. The warm coloured lines (below) correspond to transition between stable and unstable configurations, while the cold coloured lines (above) transition between solid and hollow phases. As we lower the charge q both $\psi_c$ and $\psi_h$ increase, however their difference remains roughly unchanged, highlighting how hollow solutions are unstable for all range of parameters investigated.}
\label{UnstablexHollow-compa}
\end{figure}

In order to numerically determine $\psi_c$ we fix $q$ and $m^2$ and search for the highest value of the star mass $M$ as a function of $\psi(0)$. Similarly to the phase diagrams above (figures \ref{SH3D} and \ref{SH4D}), we find, for each value of $q$, a line dividing stable and unstable regimes in the $m^2$ vs. $\psi_0$ plane. We observe that, within the range of parameters we studied, we always have $\psi_c<\psi_h$ (as seen in figures \ref{UnstablexHollow} and \ref{UnstablexHollow-compa}), indicating that the hollow solutions we found, while being perfectly fine in a mathematical sense, do not correspond to a physically preferred phase and suffer from dynamical instabilities.  


\section{Final Comments}

In this paper we numerically investigated the behaviour of extremal, codimension 2, spacelike surfaces in charged, asymptotically AdS, boson star backgrounds. Our main goal was to establish the conditions for which families of minimal area spacelike surfaces anchored on the boundary fail to fully cover the bulk of the spacetime. As discussed in the Introduction, this study was motivated by recent ideas regarding a possible connection between the holographic description of entaglement entropy and the gravity dual of a reduced density matrix as discussed in \cite{Bartek.et.al1}. 

We observed that the relation between the star's mass as well the the star's charge and the central density of the scalar field in four dimensions behave much like what is known for both neutral boson stars in AdS and charged boson stars in flat space. Notably, the existence of a maximum mass for a finite $\psi(0)=\psi_c$ strongly hints towards the presence of a stability threshold and can be used to infer the physical feasibility of the hollow solutions we were so interested in. In three dimensions we found results akin to what is known in the literature for other types of boson stars, in particular, we observed a behaviour similar as the one found in \cite{Astefanesei.Radu} for $2+1$ dimensional neutral boson stars in asymptotically AdS spacetime.

Our analysis of the behaviour of extremal surfaces with fixed boundary points led us to the conclusion that, both in three and four dimensions, for fixed $m^2$ and $q$ there is a maximum value for the central density of the scalar field $\psi_h$ for which the minimal area surfaces reach every point in the bulk space (figures \ref{SH3D} and \ref{SH4D}). Therefore one should expect that charged boson stars with a high enough $\psi(0)$ could provide a clear obstacle in the way of $w(D_A)$ being a universal candidate for $R(A)$. However we saw that, at least in four dimensions, there is good evidence indicating that solutions with $\psi(0)\ge\psi_h$ are unstable (figure \ref{UnstablexHollow}). Extrapolating well known results in the literature for both boson stars with and without gauge fields in flat space, and boson stars without gauge field in AdS space, we find that for given $m^2$ and $q$ there is a threshold central density value $\psi_c$ for which the solutions cease to be stable if $\psi(0)>\psi_c$. Remarkably, in the four dimension case in question, we found that for every pair of $m^2$ and $q$, $\psi_c<\psi_h$, i.e.: solutions for which $w(D_B)$ fail to cover the entire bulk and, in particular, $z(D_A) \supset w(D_A)$ while $w(D_A)\nsupseteq z(D_A)$, are physically unstable. Unfortunately, to the extent of this author's knowledge, much less in known about the stability of three dimensional boson stars, therefore precluding us from saying anything about the stability of both regimes we found.

We believe the results found in this work support some of the ideas discussed in \cite{Bartek.et.al1} and further explored in \cite{Czech:2012be,Bousso:2012mh,Wall}. The unstable character of hollowed solutions strengthens the proposal of $w(D_A)$ as a good candidate for $R(A)$ and complements other recent works on the subject. We also believe that a deeper understanding of extremal surfaces on charged boson stars backgrounds can serve as a fruitful test ground for numerous holographic ideas including, but not restricted to, the holographic entanglement entropy, the holographic dual of a density matrix, zero temperature quantum phase transitions \cite{CBS-phaseT}, etc.

\section*{Acknowledgements}\label{acknow}

The author would like to thank Mark Van Raamsdonk and Joanna Karczmarek for helpful discussions and comments throughout this project, and Jared Stang and Connor Behan for helpful comments on the manuscript.

\bibliographystyle{naturemag}
\bibliography{CBSref}

\begin{thebibliography}{10}
\expandafter\ifx\csname url\endcsname\relax
  \def\url#1{\texttt{#1}}\fi
\expandafter\ifx\csname urlprefix\endcsname\relax\def\urlprefix{URL }\fi
\providecommand{\bibinfo}[2]{#2}
\providecommand{\eprint}[2][]{\url{#2}}

\bibitem{thooft}
\bibinfo{author}{'t~Hooft, G.}
\newblock \bibinfo{title}{{Dimensional reduction in quantum gravity.}}
\newblock \emph{\bibinfo{journal}{THU-93-26}}  (\bibinfo{year}{1993}).
\newblock \eprint{9310026}.

\bibitem{susskind}
\bibinfo{author}{Susskind, L.}
\newblock \bibinfo{title}{{The World as a hologram.}}
\newblock \emph{\bibinfo{journal}{J.Math.Phys.}} \textbf{\bibinfo{volume}{36}},
  \bibinfo{pages}{6377--6396} (\bibinfo{year}{1994}).
\newblock \eprint{9409089}.

\bibitem{Maldacena:1997re}
\bibinfo{author}{Maldacena, J.~M.}
\newblock \bibinfo{title}{{The large N limit of superconformal field theories
  and supergravity}}.
\newblock \emph{\bibinfo{journal}{Adv. Theor. Math. Phys.}}
  \textbf{\bibinfo{volume}{2}}, \bibinfo{pages}{231--252}
  (\bibinfo{year}{1998}).
\newblock \eprint{hep-th/9711200}.

\bibitem{Witten:1998qj}
\bibinfo{author}{Witten, E.}
\newblock \bibinfo{title}{{Anti-de Sitter space and holography}}.
\newblock \emph{\bibinfo{journal}{Adv. Theor. Math. Phys.}}
  \textbf{\bibinfo{volume}{2}}, \bibinfo{pages}{253--291}
  (\bibinfo{year}{1998}).
\newblock \eprint{hep-th/9802150}.

\bibitem{McGreevy:2009xe}
\bibinfo{author}{McGreevy, J.}
\newblock \bibinfo{title}{{Holographic duality with a view toward many-body
  physics}}.
\newblock \emph{\bibinfo{journal}{Adv.High Energy Phys.}}
  \textbf{\bibinfo{volume}{2010}}, \bibinfo{pages}{723105}
  (\bibinfo{year}{2010}).
\newblock \eprint{0909.0518}.

\bibitem{Hartnoll:2009sz}
\bibinfo{author}{Hartnoll, S.~A.}
\newblock \bibinfo{title}{{Lectures on holographic methods for condensed matter
  physics}}.
\newblock \emph{\bibinfo{journal}{Class.Quant.Grav.}}
  \textbf{\bibinfo{volume}{26}}, \bibinfo{pages}{224002}
  (\bibinfo{year}{2009}).
\newblock \eprint{0903.3246}.

\bibitem{ryu.takayanagi}
\bibinfo{author}{Ryu, S.} \& \bibinfo{author}{Takayanagi, T.}
\newblock \bibinfo{title}{{Holographic derivation of entanglement entropy from
  AdS/CFT.}}
\newblock \emph{\bibinfo{journal}{Phys.Rev.Lett.}}
  \textbf{\bibinfo{volume}{96}} (\bibinfo{year}{2006}).
\newblock \eprint{0603001}.

\bibitem{Bartek.et.al1}
\bibinfo{author}{Czech, B.}, \bibinfo{author}{Karczmarek, J.~L.},
  \bibinfo{author}{Nogueira, F.} \& \bibinfo{author}{Van~Raamsdonk, M.}
\newblock \bibinfo{title}{{The Gravity Dual of a Density Matrix}}.
\newblock \emph{\bibinfo{journal}{Class.Quant.Grav.}}
  \textbf{\bibinfo{volume}{29}} (\bibinfo{year}{2012}).
\newblock \eprint{1204.1330}.

\bibitem{Wall}
\bibinfo{author}{Wall, A.~C.}
\newblock \bibinfo{title}{{Maximin Surfaces, and the Strong Subadditivity of
  the Covariant Holographic Entanglement Entropy}}  (\bibinfo{year}{2012}).
\newblock \eprint{1211.3494}.

\bibitem{kaup}
\bibinfo{author}{Kaup, D.~J.}
\newblock \bibinfo{title}{{Klein-Gordon Geon.}}
\newblock \emph{\bibinfo{journal}{Phys.Rev.}} \textbf{\bibinfo{volume}{172}},
  \bibinfo{pages}{1331Ð1342} (\bibinfo{year}{1968}).

\bibitem{Ruffini.Bonazzola}
\bibinfo{author}{Ruffini, R.} \& \bibinfo{author}{Bonazzola, S.}
\newblock \bibinfo{title}{{Systems of self gravitating particles in general
  relativity and the concept of an equation of state}}.
\newblock \emph{\bibinfo{journal}{Phys.Rev.}} \textbf{\bibinfo{volume}{187}},
  \bibinfo{pages}{1767Ð1783} (\bibinfo{year}{1969}).

\bibitem{Schunck:2003kk}
\bibinfo{author}{Schunck, F.} \& \bibinfo{author}{Mielke, E.}
\newblock \bibinfo{title}{{General relativistic boson stars}}.
\newblock \emph{\bibinfo{journal}{Class.Quant.Grav.}}
  \textbf{\bibinfo{volume}{20}}, \bibinfo{pages}{R301--R356}
  (\bibinfo{year}{2003}).
\newblock \eprint{0801.0307}.

\bibitem{Jetzer:1991jr}
\bibinfo{author}{Jetzer, P.}
\newblock \bibinfo{title}{{Boson stars}}.
\newblock \emph{\bibinfo{journal}{Phys.Rept.}} \textbf{\bibinfo{volume}{220}},
  \bibinfo{pages}{163--227} (\bibinfo{year}{1992}).

\bibitem{Liebling:2012fv}
\bibinfo{author}{Liebling, S.~L.} \& \bibinfo{author}{Palenzuela, C.}
\newblock \bibinfo{title}{{Dynamical Boson Stars}}.
\newblock \emph{\bibinfo{journal}{Living Rev.Rel.}}
  \textbf{\bibinfo{volume}{15}}, \bibinfo{pages}{6} (\bibinfo{year}{2012}).
\newblock \eprint{1202.5809}.

\bibitem{Liddle:1993ha}
\bibinfo{author}{Liddle, A.~R.} \& \bibinfo{author}{Madsen, M.~S.}
\newblock \bibinfo{title}{{The Structure and formation of boson stars}}.
\newblock \emph{\bibinfo{journal}{Int.J.Mod.Phys.}}
  \textbf{\bibinfo{volume}{D1}}, \bibinfo{pages}{101--144}
  (\bibinfo{year}{1992}).

\bibitem{Astefanesei.Radu}
\bibinfo{author}{Astefanesei, D.} \& \bibinfo{author}{Radu, E.}
\newblock \bibinfo{title}{{Boson stars with negative cosmological constant}}.
\newblock \emph{\bibinfo{journal}{Nucl.Phys.}} \textbf{\bibinfo{volume}{B665}},
  \bibinfo{pages}{594--622} (\bibinfo{year}{2003}).
\newblock \eprint{gr-qc/0309131}.

\bibitem{CBS-phaseT}
\bibinfo{author}{Hu, S.}, \bibinfo{author}{Liu, J.~T.} \&
  \bibinfo{author}{Zayas, L. A.~P.}
\newblock \bibinfo{title}{{Charged Boson Stars in AdS and a Zero Temperature
  Phase Transition.}}  (\bibinfo{year}{2012}).
\newblock \eprint{1209.2378}.

\bibitem{Gentle:2011kv}
\bibinfo{author}{Gentle, S.~A.}, \bibinfo{author}{Rangamani, M.} \&
  \bibinfo{author}{Withers, B.}
\newblock \bibinfo{title}{{A Soliton Menagerie in AdS}}.
\newblock \emph{\bibinfo{journal}{JHEP}} \textbf{\bibinfo{volume}{1205}},
  \bibinfo{pages}{106} (\bibinfo{year}{2012}).
\newblock \eprint{1112.3979}.

\bibitem{Jetzer:1989us}
\bibinfo{author}{Jetzer, P.}
\newblock \bibinfo{title}{{Stability of charged boson stars}}.
\newblock \emph{\bibinfo{journal}{Phys.Lett.}} \textbf{\bibinfo{volume}{B231}},
  \bibinfo{pages}{433} (\bibinfo{year}{1989}).

\bibitem{Sakamoto:1998hq}
\bibinfo{author}{Sakamoto, K.} \& \bibinfo{author}{Shiraishi, K.}
\newblock \bibinfo{title}{{Boson stars with large selfinteraction in
  (2+1)-dimensions: An Exact solution}}.
\newblock \emph{\bibinfo{journal}{JHEP}} \textbf{\bibinfo{volume}{9807}},
  \bibinfo{pages}{015} (\bibinfo{year}{1998}).
\newblock \eprint{gr-qc/9804067}.

\bibitem{Sakamoto:1998aj}
\bibinfo{author}{Sakamoto, K.} \& \bibinfo{author}{Shiraishi, K.}
\newblock \bibinfo{title}{{Exact solutions for boson fermion stars in
  (2+1)-dimensions}}.
\newblock \emph{\bibinfo{journal}{Phys.Rev.}} \textbf{\bibinfo{volume}{D58}},
  \bibinfo{pages}{124017} (\bibinfo{year}{1998}).
\newblock \eprint{gr-qc/9806040}.

\bibitem{Hamilton:2005ju}
\bibinfo{author}{Hamilton, A.}, \bibinfo{author}{Kabat, D.~N.},
  \bibinfo{author}{Lifschytz, G.} \& \bibinfo{author}{Lowe, D.~A.}
\newblock \bibinfo{title}{{Local bulk operators in AdS/CFT: A Boundary view of
  horizons and locality}}.
\newblock \emph{\bibinfo{journal}{Phys.Rev.}} \textbf{\bibinfo{volume}{D73}},
  \bibinfo{pages}{086003} (\bibinfo{year}{2006}).
\newblock \eprint{hep-th/0506118}.

\bibitem{Hamilton:2006az}
\bibinfo{author}{Hamilton, A.}, \bibinfo{author}{Kabat, D.~N.},
  \bibinfo{author}{Lifschytz, G.} \& \bibinfo{author}{Lowe, D.~A.}
\newblock \bibinfo{title}{{Holographic representation of local bulk
  operators}}.
\newblock \emph{\bibinfo{journal}{Phys.Rev.}} \textbf{\bibinfo{volume}{D74}},
  \bibinfo{pages}{066009} (\bibinfo{year}{2006}).
\newblock \eprint{hep-th/0606141}.

\bibitem{Heemskerk:2012mn}
\bibinfo{author}{Heemskerk, I.}, \bibinfo{author}{Marolf, D.},
  \bibinfo{author}{Polchinski, J.} \& \bibinfo{author}{Sully, J.}
\newblock \bibinfo{title}{{Bulk and Transhorizon Measurements in AdS/CFT}}.
\newblock \emph{\bibinfo{journal}{JHEP}} \textbf{\bibinfo{volume}{1210}},
  \bibinfo{pages}{165} (\bibinfo{year}{2012}).
\newblock \eprint{1201.3664}.

\bibitem{Kabat:2011rz}
\bibinfo{author}{Kabat, D.}, \bibinfo{author}{Lifschytz, G.} \&
  \bibinfo{author}{Lowe, D.~A.}
\newblock \bibinfo{title}{{Constructing local bulk observables in interacting
  AdS/CFT}}.
\newblock \emph{\bibinfo{journal}{Phys.Rev.}} \textbf{\bibinfo{volume}{D83}},
  \bibinfo{pages}{106009} (\bibinfo{year}{2011}).
\newblock \eprint{1102.2910}.

\bibitem{Bousso:2012mh}
\bibinfo{author}{Bousso, R.}, \bibinfo{author}{Freivogel, B.},
  \bibinfo{author}{Leichenauer, S.}, \bibinfo{author}{Rosenhaus, V.} \&
  \bibinfo{author}{Zukowski, C.}
\newblock \bibinfo{title}{{Null Geodesics, Local CFT Operators and AdS/CFT for
  Subregions}}  (\bibinfo{year}{2012}).
\newblock \eprint{1209.4641}.

\bibitem{Czech:2012be}
\bibinfo{author}{Czech, B.}, \bibinfo{author}{Karczmarek, J.~L.},
  \bibinfo{author}{Nogueira, F.} \& \bibinfo{author}{Van~Raamsdonk, M.}
\newblock \bibinfo{title}{{Rindler Quantum Gravity}}.
\newblock \emph{\bibinfo{journal}{Class.Quant.Grav.}}
  \textbf{\bibinfo{volume}{29}}, \bibinfo{pages}{235025}
  (\bibinfo{year}{2012}).
\newblock \eprint{1206.1323}.

\bibitem{Hubeny}
\bibinfo{author}{Hubeny, V.~E.}
\newblock \bibinfo{title}{{Extremal surfaces as bulk probes in AdS/CFT.}}
\newblock \emph{\bibinfo{journal}{JHEP}} \textbf{\bibinfo{volume}{093}}
  (\bibinfo{year}{2012}).
\newblock \eprint{1203.1044}.

\end{thebibliography}

\end{document}